\UseRawInputEncoding
\documentclass[preprint,12pt]{elsarticle}
\usepackage{listings}
\usepackage[utf8]{inputenc}  
\usepackage{listings}
\usepackage{xcolor}
\usepackage[colorlinks=true, linkcolor=blue, citecolor=blue, urlcolor=blue]{hyperref}

\usepackage{ragged2e}

\usepackage{graphicx}
\usepackage{dcolumn}
\usepackage{xcolor}
\usepackage{bm}
\usepackage{subcaption}

\usepackage{calligra}
\usepackage[T1]{fontenc}
\usepackage{egothic}
\usepackage[T1]{fontenc}
\newfont{\rsfsten}{rsfs10 scaled 1200}
\newfont{\rsfsseven}{rsfs10 scaled 1200}
\newfont{\rsfsfive}{rsfs10 scaled 1200}
\usepackage{epsfig}
\usepackage{units}
\usepackage[utf8]{inputenc}
\usepackage{hyperref}
\usepackage{tabularx}

\newcommand{\be}{\begin{equation}}
\newcommand{\ee}{\end{equation}}
\newcommand{\bea}{\begin{eqnarray}}
\newcommand{\eea}{\end{eqnarray}}

\usepackage{comment}

\usepackage{amssymb}
\usepackage{amsmath}

\journal{Chemical Physics }

\begin{document}

\begin{frontmatter}


\title{Mutation in DNA: A Quantum Mechanical Non-Adiabatic Model}

\author{Hossien Hossieni}
\ead{hossien.hossieni@univsul.edu.iq}

\affiliation{
  organization={Physics Department, College of Science, University of Sulaimani},
  city={Sulaimani},
  state={Kurdistan Region},
  postcode={46001},
  country={Iraq}
}



\author{------------} 


\begin{abstract}
We propose a new analytical potential function to model proton transfer in the adenine-thymine base pair and develop a non-adiabatic quantum mechanical framework to calculate genetic mutation probabilities. This potential has been used to calculate the probability of mutation in a non-adiabatic process. The results of the new model have been shown to be consistent with the findings of other researchers.
\end{abstract}




\end{frontmatter}

\section{Introduction}
In biology,two DNA strands are connected by hydrogen bonds. There are four nitrogenous bases types in DNA: cytosine (C), guanine (G), adenine (A), and thymine (T). The ordering of different nitrogenous bases holds different genetic information. This structure gives the DNA 
the ability to make an organism that can replicate itself. Occasionally, 
errors occur in the arrangement of nitrogenous bases along the strands during DNA replication. Typically, the DNA that has errors will be destroyed. However, in rare cases, these errors can persist, allowing replication of the defective DNA. In this way, the information encoded in chains of nitrogenous bases  evolves and increases by complexity. When mutations happen in the coding region of the genes, they can cause destructive results. On average, DNA polymerase introduces one error for every $10000-100000$ bases during replication \cite{prestona2010}. If such errors occur in coding regions, they are transcribed into RNA and translated into proteins, potentially altering protein function.

The cause for mutations is assumed to originate at the molecular level. According to the classical views, heredity transfers mutations from one generation to the next\cite{KlugGenetics2017}. At the cellular level, cellular machines probably can make an error in replicating or transferring DNA chains,resulting in mutation. 
Although the cells have some mechanism to correct these errors, the DNA's helix is  flexible and some errors survive, meaning that the double helix can maintain certain structural defects \cite{Xu2019}.
The defects originate from quantum shifts (tautomeric transitions) of the nucleotide bases. If DNA polymerase fails to correct errors in the nucleotide sequence, mistaken bases 
are incorporated into RNA, leading to mutations. The rate of mutation  by this mechanism is $ 1/100000$ nucleotides\cite{Carey2015}. Watson and Crick suggested that genetic codes are saved in the form of hydrogen bonds inside nucleotide bases of $A-T$ and $C-G$,  forming a sequence of the double structure of the DNA molecule.
A chemical process that interconverts isomers is known as tautomerization. This process often involves the movement of a hydrogen ion 
and the rearrangement of electron bonds,  typically between a keto and enol forms.  
A hydrogen bond is an electrostatic interaction between a hydrogen atom that is covalently bonded to an electronegative atom, such as oxygen or nitrogen, and another nearby electronegative atom. A proton (the hydrogen nucleus) participates in this interaction. During DNA replication, when the double helix begins to unwind, each strand serves as a template to form its complementary nucleotides, resulting in two DNA molecules, each containing the original genetic information. 
There are some tautomers for hydrogen bonds, so that some double bases may have been made in the wrong way. Lowdin \cite{Lowdin1963} showed the cause behind hydrogen bond tautomerization and, consequently, the genetic mutation is probably quantum tunneling. In the Lowdin model, the proton of the hydrogen bond is trapped in an asymmetric double potential well between the adenine and thymine bases. The proton can move from one well to another with a higher energy minimum. Several studies have been performed to calculate quantum tunneling effects.

Several studies \cite{Lowdin1963,Slocombe2021,Srivastava2019,Brovarets2015} mentioned that for tunneling, the proton should absorb some energy to move to higher states. The researchers show lifetime of  $A^*-T^*$  is on the order of $ 10^{-14}$ second, and this is a very short time to survive in the replication machinery. Borovets and Hovorun \cite{Brovarets2015} did another enlightenment research. They also considered the effect of medium. They concluded that tautomerization could not cause a genetic mutation in DNA replication. In 2015, Godbeer et al. \cite{Godbeer2015} proposed a new model for quantum tunneling in double bases of adenine-thymine. 
They concluded, based on their model, that the probability due to tunneling in the presence of a medium is only $2\times 10^{-9} $. Of course, that is a low ratio. Nevertheless, Bovarets and Hovorun, who have done several pieces of research in this field, believe that quantum tunneling occurs in double-mistake connections such as the nucleobase mispairs:

\begin{center}
\[
\begin{aligned}
H \cdot H^{*} &\leftrightarrow H \cdot H^{*},\\
T \cdot T^{*} &\leftrightarrow T \cdot T^{*},\\
G \cdot G^{*} &\leftrightarrow G \cdot G^{*},\\
C \cdot C^{*} &\leftrightarrow C \cdot C^{*},\\
A \cdot A^{*} &\leftrightarrow A \cdot A^{*}.
\end{aligned}
\]
\end{center}
A compact notation, $X \cdot X^* \leftrightarrow X \cdot X^*$, can be used to represent femtosecond-scale interconversion for all identical-base pairs, including $G \cdot G^*$, $C \cdot C^*$, $A \cdot A^*$, in addition to $T \cdot T^*$ and $H \cdot H^*$. Here, $X$ can be any of T, H, G, C, or A. This notation indicates participation of two identical bases in a hydrogen-bonded configuration that allows the proton to shuttle back and forth, converting the canonical form of $X$ to its rare tautomer $X^*$. Such a double-well proton-exchange system is predicted for all the above transient mispairs \cite{Slocombe2022}.

DNA can be damaged through a process known as photoinduced electron transfer (PET) mechanism. In this mechanism, an incoming photon excites an electron, which can then interact with a proton or another part of the molecular structure, raising its energy and potentially disrupting chemical bonds~\cite{Christopher2018,Haiharan2010}. Ultraviolet (UV) radiation from the Sun, for example, can trigger such interactions, breaking chemical bonds within DNA and leading to mutations. Individual nucleotide bases have excited states that typically last only a few picoseconds and usually decay rapidly, minimizing the likelihood of harmful chemical reactions. However, excited states that involve both strands of DNA can persist for longer periods.Beckstead~\cite{Beckstead2017} showed that double radical ions can form through photoinduced electron transfer within nucleotide bases, with lifetimes on the order of hundreds of picoseconds. Furthermore, picosecond-scale excited-state lifetimes have also been observed in double-stranded DNA~\cite{Duchi2019}.

\section{Adenine-Thymine Potential}
A large number of studies on proton transfer during tautomerization in hydrogen bonds have been conducted without explicitly considering quantum tunneling. Many of these approaches rely on \textit{ab initio} calculations and potential energy surfaces~\cite{Florian1994, Scheiner1991}, while others use soliton-based models~\cite{Pnevmatikos1991, Kryachko1992}.

Ref.\cite{Godbeer2015} studied the energies of canonical and tautomeric adenine--thymine base pairs, $A\!-\!T$ and $A^{*}\!-\!T^{*}$, respectively, and constructed the corresponding potential energy surface using functional density theory. They employed a generalized Liouville equation, which included an additional term to account for dissipation:

\begin{equation}\label{ch4eq1}
i\hbar(\frac{d\rho}{dt})=[{\bf H}{\bf ,} {\bf \rho}]+ {\bf {L\rho}} 
\end{equation}
where the extra term is written in the form

\begin{equation}\label{ch4eq2}
\bf {L\rho}=\sum_{ij}^{ }(\bf {A_{ij}\rho}A_{ij}^\dagger) -\frac{1}{2}\{\bf{A_{ij}^\dagger A_{ij}} , \bf{\rho}\} )
\end{equation}

with 
\begin{equation}\label{ch4eq3}
\bf {A_{ij}}= \sqrt{W_{ij}}|\phi_{i} \rangle \langle \phi_{j}|
\end{equation}
    
$W_{ij}$ are  environment-induced transition rates between well states $|\phi_i\rangle$ and $\langle\phi_j|$.

The energy levels of the $A\!-\!T$ potential, along with the probabilities of finding the proton in the right-hand well, were determined using quantum tunneling techniques.

\begin{figure}[h!]
    \centering
    \includegraphics[width=1\textwidth]{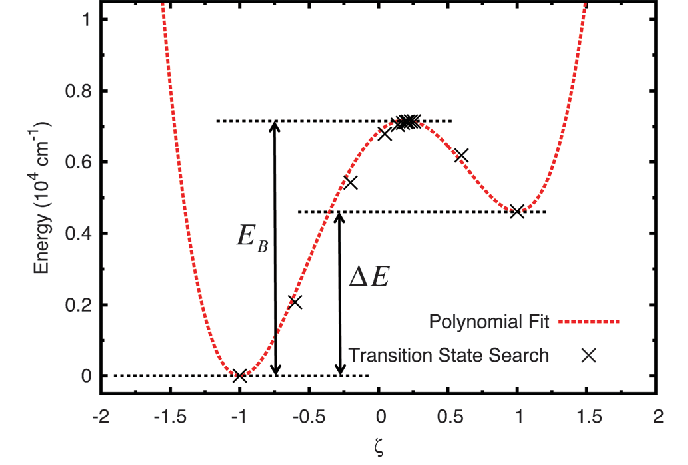} 
    \caption{The Adenine-Thymine potential curve from Eq.\eqref{ch3eq4} shows an asymmetric double-well potential, indicating the energy difference between A--T and its tautomer $A^{*} - T^{*}$. At the bottom of the wells, the proton is stable but at different energy levels. The central barrier is slightly shifted from $\zeta = 0$, reflecting the inequality in energy levels. The curve is scaled from $eV$ to cm$^{-1}$ using a factor of 8065.73. Reused from \cite{Godbeer2015}. Licensed under a CC-BY-3.0 License.
    }
    \label{ch3f1}
\end{figure}

There are several models for the adenine–thymine potential, all of which agree that it has the form of an asymmetric double well~\cite{Bountis1991}. Ref.~\cite{Godbeer2015}, using the canonical A–T structure from the S22 database calculated at the MP2/cc-pVTZ level, pointed out that the tautomeric form $A^{*}\!-\!T^{*}$ is not included in the database and therefore its structure had to be estimated before geometric optimization. To do this, the optimized B3LYP geometry of A–T was modified to generate an initial geometry for $A^{}!-!T^{}$, and a fitting procedure was applied to obtain the following potential:

\begin{equation}\label{ch3eq4}
V(\zeta)=8065.73\cdot (-4464.49+0.429\zeta-1.126\zeta^2-0.143\zeta^3+0.563\zeta^4)
\end{equation}
Here, $V$ is expressed in $\mathrm{cm^{-1}}$, and $\zeta$ is the dimensionless proton coordinate along the $x$ axis, defined as $\zeta = x/a$. The two minima are separated by $2a$, located at $\zeta = \pm 1$, while the central maximum is slightly shifted from $\zeta = 0$ due to the potential’s asymmetry. The prefactor $8065.73$ converts the potential from eV to $\mathrm{cm^{-1}}$. The asymmetric double-well potential is plotted in Fig.~\ref{ch3f1}.

\begin{figure}[h!]
    \centering
    \includegraphics[width=1\textwidth]{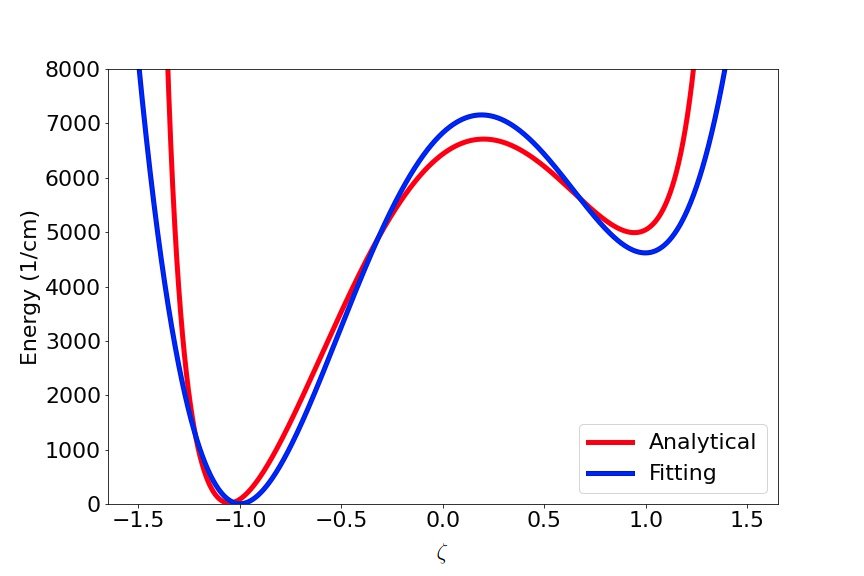}
    \caption[Chapter 3: Fitting and analytical potential in a figure.]{
        Variation of the adenine–thymine potential obtained analytically~\cite{Sitnitsky2017} and by fitting~\cite{Godbeer2015}. The symbol of the horizontal axis represents both $\zeta$ and $x$ in Eqs. \eqref{ch3eq4}and \eqref{ch3eq5}, respectively. We set  the values of the parameters in  Eq.\eqref{ch3eq5} $h = 70$, $a = 300$, $b = 850$ and $c=\sqrt{h}$.
    }
    \label{ch3f2}
\end{figure}

Ref.\cite{Sitnitsky2017} has presented an analytical solution for the hydrogen bond in  Potassium bicarbonate $\rm KHCO_{3}$ that has a potential similar to the exact  formula:
\begin{equation}\label{ch3eq5}
V(x)=h\tan^2x+a\sin x- b\sin^2x+c\frac{\sin x}{\cos^2x}
\end{equation}
where $h$ and $c$ are parameters related to the barrier width and the asymmetry of the potential well, and $a$ and $b$ are  the asymmetry and the barrier height parameters, respectively. The two potentials in Eqs. \eqref{ch3eq4} and \eqref{ch3eq5} were plotted for comparison in Fig. \ref{ch3f2}. Overall, we observe good agreement between the fitted model of Ref. \cite{Godbeer2015} and the analytical solution of Ref. \cite{Sitnitsky2017}.

For the one-dimensional motion of a proton the time-independent Schrodinger equation is:
 \begin{equation}\label{ch3eq6}
 \left[ \frac{\hbar^2}{2m}\frac{d^2}{d\zeta^2}+(E-V(\zeta))\right] \psi=0 
 \end{equation}
 
\begin{figure}[htb!] 
    \centering
    \includegraphics[width=1\textwidth]{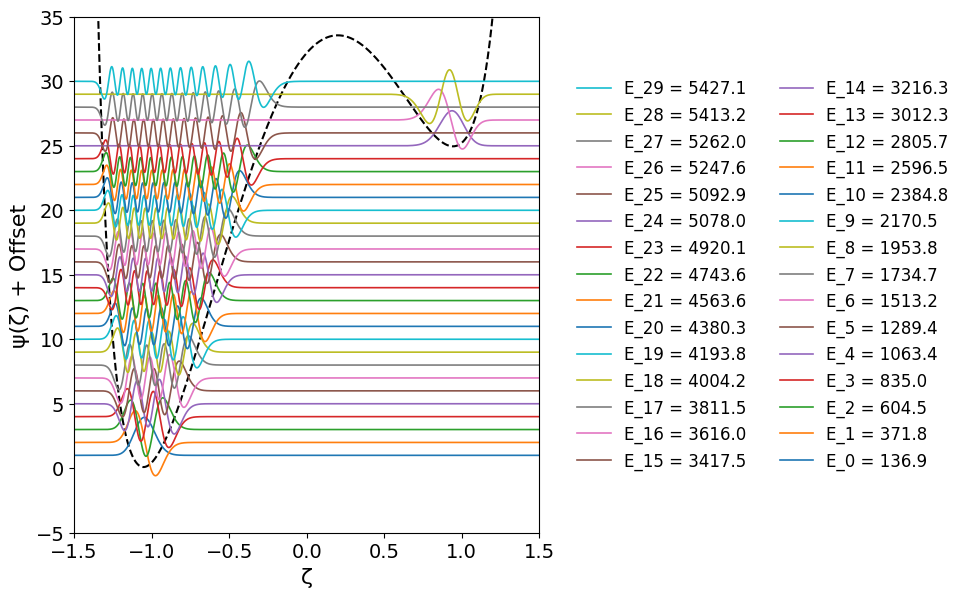} 
    \caption[The first 30 wave functions]{
     {The first  thirty eigenstates of the analytical potential in Eq.\eqref{ch3eq5} are shown, along with their corresponding energies $E_i \ (i=0,1,2...29)$ in cm$^{-1}$. The double-well potential is included to illustrate the spatial distribution of these states. These eigenstates are directly related to proton transfer, which is a key step in tautomeric mutations.}}
    \label{ch3f3}
\end{figure}

In the present work we solved Eq.\eqref{ch3eq6} numerically with the potential $V(\zeta)$ of Eq.\eqref{ch3eq5} to obtain all eigenstates and eigenenergies of the proton. For this purpose, a Python code was developed and made publicly available\footnote{\url{https://github.com/hossienhossieni-bot/MUTATION2}}. The first 30 eigenstates obtained are displayed in Fig. \ref{ch3f3}.

The basic condition for a mutation to occur is that the particle should be on the right-hand side of the potential ($\zeta>0$). After the $24^{\rm th}$ eigenstate, penetration to the right-hand side occurs, and some eigenstates appear in the second well (see Fig.~\ref{ch3f3}). It is clear that the energies $E_{24} = 5078\ \mathrm{cm^{-1}}$ and $E_{25} = 5092.9\ \mathrm{cm^{-1}}$ are lower than the height of the potential barrier between the two wells (greater than $7000\ \mathrm{cm^{-1}}$; see the peak in Fig.~\ref{ch3f1}). This means that the proton has tunneled to the second well.

\section{A Non-Adiabatic Quantum Mechanical Model for DNA Mutation}

Nonadiabatic dynamics in molecules involve processes in which the motion of a nucleus is affected by more than one electronic state. These processes can occur when a molecule is irradiated with light. In such a situation, whenever two or more states have similar energies and the state-to-state couplings are large, population transfer will occur from one state to another \cite {Mai2018}.

Fig.~\ref{ch3f2} shows that the peak of the potential energy in the present model is in good agreement with the average energy of the hydrogen bond, approximately 6300~$\mathrm{cm^{-1}}$~\cite{Steiner2002}. This energy lies within the infrared (IR) spectral range, indicating that IR absorption can activate the proton and potentially trigger nonadiabatic transitions \cite{Andreas2007}.

A molecule is activated to an excited electronic state as a result of the absorption of IR. After excitation, it deactivates and returns to the ground state through different relaxation mechanisms, such as radiative emission via fluorescence and/or phosphorescence, or non-radiative pathways such as internal conversion (heat dissipation)\cite{Ibele2020}. The fastest mechanism is internal conversion, which can occur in a few picoseconds. For example, UV photoexcited nucleic acid bases deactivate around a picosecond \cite {Barbatti2008,Barbatti2010}.
On the scale of the biological process, this is a very short time.\\

Simple potential energy models cannot represent the complexity of the potential energy of the excited state. The potential energy of an electronically excited state is often very close to that of other excited states. Consequently, during the relaxation (a nonadiabatic process), the molecule may jump to other states. The relaxation of electronically excited molecules is fundamentally studied using nonadiabatic dynamics\cite{Barbatti2020}.\\
 However,
Photoexcitation, thermal fluctuations, and molecular vibrations can transiently reshape the potential-energy landscape but most important  internal mechanism is Hiesenberg uncertainty  principle. Zewail’s femtochemistry experiments demonstrated ultrafast barrier switching \cite{Zewail2000}. These observations justify our model of a sudden, non-adiabatic barrier reduction.

Using the Heisenberg energy–time uncertainty relation

 \begin{equation}\label{ch4eq7}
\Delta E\,\Delta t \approx \hbar
 \end{equation}
 the lifetime of the transient tautomer is \cite{Slocombe2022}
 \begin{equation}\label{ch4eq8}
\Delta t = t_{1/2} = \frac{\ln 2}{k_r} \approx 4.1 \times 10^{-14}\,\text{s}  \approx 41\,\text{fs},
 \end{equation}

where \(t_{1/2}\) is the half-life of the tautomer and \(k_r\) is the reverse rate,
\(k_r = 1.69 \times 10^{13}\,\text{s}^{-1}\).
This result shows that the tautomer is extremely short-lived.
we obtained a natural timescale of $10^{-15}$\,s, consistent with ultrafast tautomerization.
This is the minimum time during which the proton almost cannot sense the barrier and can move nearly freely from the left side of the well to the right, potentially causing a misconnection during DNA replication. Within this short interval, the process is non-adiabatic, and the interaction with the environment is negligible. 

In this work, we consider a nonadiabatic model, assuming that the peak of energy produced by the interaction of a photon with the system disappears suddenly during the relaxation time (Fig. \ref{ch3f4}). Consequently, a new potential energy governs the process over a very short time interval. Since $b$ is the barrier height parameter of the potential given by Eq.~\eqref{ch3eq5}, this new potential can be obtained by setting $b=0$:

 \begin{equation}\label{ch3eq9}
V^{nonad}(\zeta)=h\tan^2\zeta+a\sin \zeta+c\frac{\sin \zeta}{\cos^2 \zeta}
\end{equation}

\begin{figure}[h!]
 \centering
 \includegraphics[width=1\textwidth]{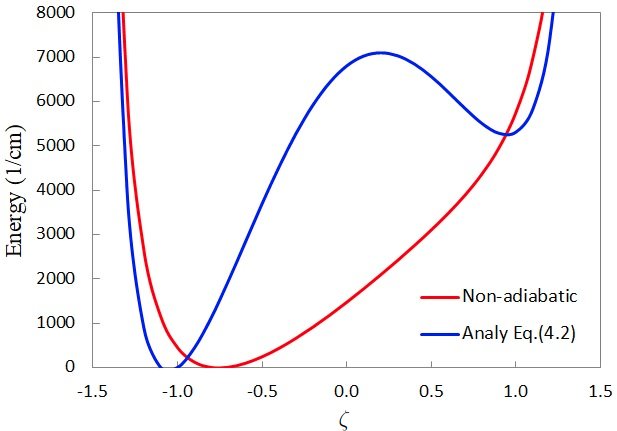}                     
\caption[\large{Chapter 3: Non-Adiabatic vs Analytic Potential}]{
Comparison of the non-adiabatic potential (\eqref{ch3eq9}) with the analytic double-well potential (Eq.~\eqref{ch3eq4}).
}  
\label{ch3f4}

 \label{ch3f4}                                                         
\end{figure} 

\begin{figure}[h!]
 \centering
 \includegraphics[width=1\textwidth]{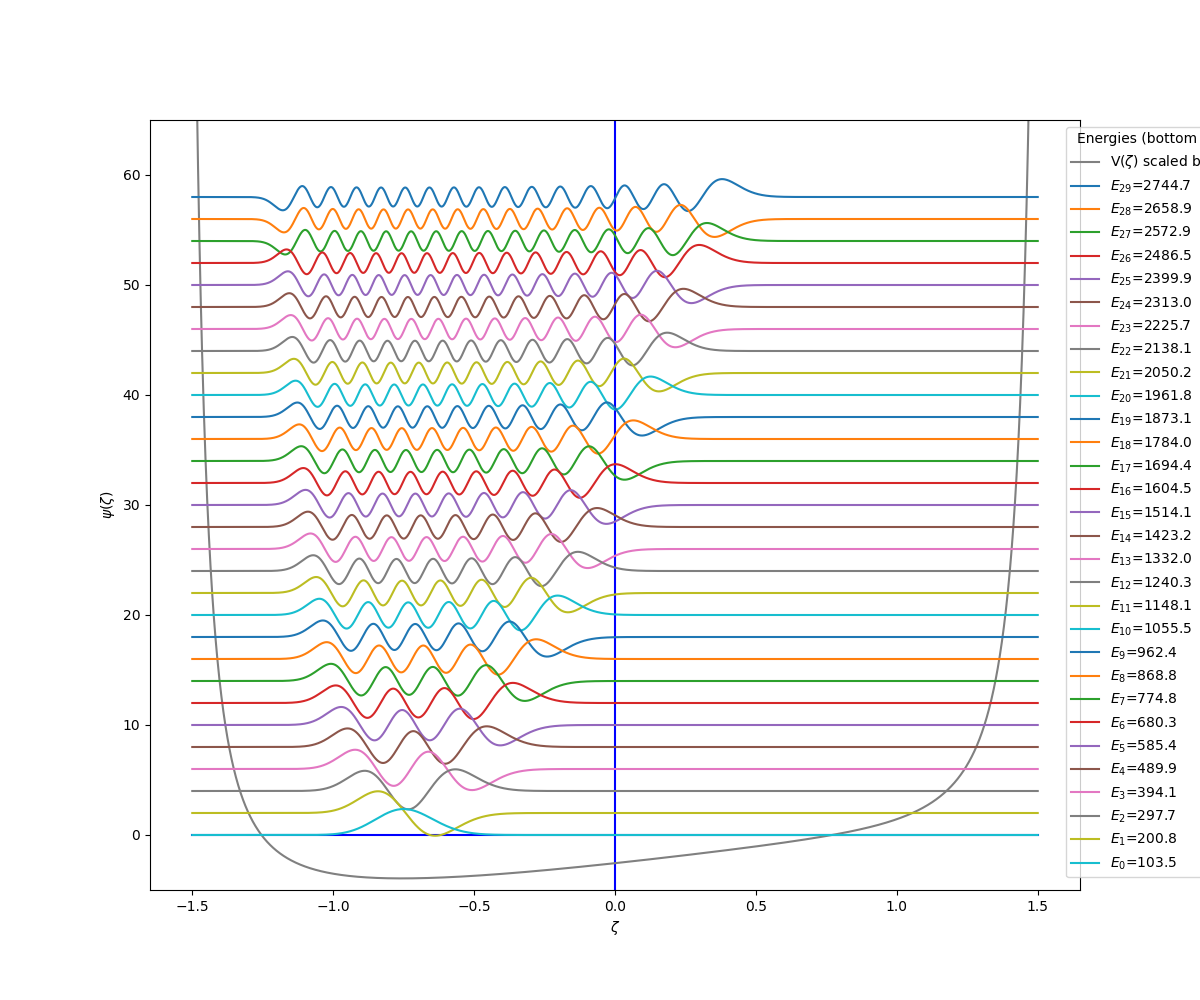}            
 \caption[{Chapter 3: Three wave functions during non-adiabatic process.}]{{The first thirty eigenstates for the non-adiabatic process along with their energies $E_i (i=0,1,2,...29)$ in $cm^{-1}$. The non-adiabatic potential is also shown. }}  
 \label{ch3f5}                                                         
\end{figure}

By solving the time-independent Schrödinger  Eq.~\eqref{ch3eq6} using the proposed potential $V(\zeta)$ from Eq.~\eqref{ch3eq9}, the eigenenergies and eigenstates were obtained. The first thirty eigenstates are plotted in Fig.\ref{ch3f5}.

To quantify the population transfer between the double-well and single-well states, we expand the ground-state wave function of the proton in the double-well, $\psi_0$, in terms of the eigenstates of the single-well potential, $\phi_n$:

 \begin{equation}\label{ch4eq5}
 \psi_0= \sum_{n=0}^{\infty}c_n\phi_n
 \end{equation}
where the expansion coefficients are

 \begin{equation}\label{ch4eq6}
  c_n=\int_{-\infty}^{+\infty} \phi_n^*\psi_0 d\zeta
 \end{equation}

The probability of finding the proton in the $\phi_n$ eigenstate is given by $|c_n|^2$. Fig. \ref{ch3f8}  shows the calculated probabilities of finding the proton in each of the first 25 eigenstates, namely, $\phi_0$ to $\phi_{24}$. Fig. \ref{ch3f8} reveals that the probability increases until the $5^{th}$ state and decreases for higher eigenstates up to the $15^{th}$, above which it becomes very small.
 
 \begin{figure}[h!]
  \centering
  \includegraphics[width=1\textwidth]{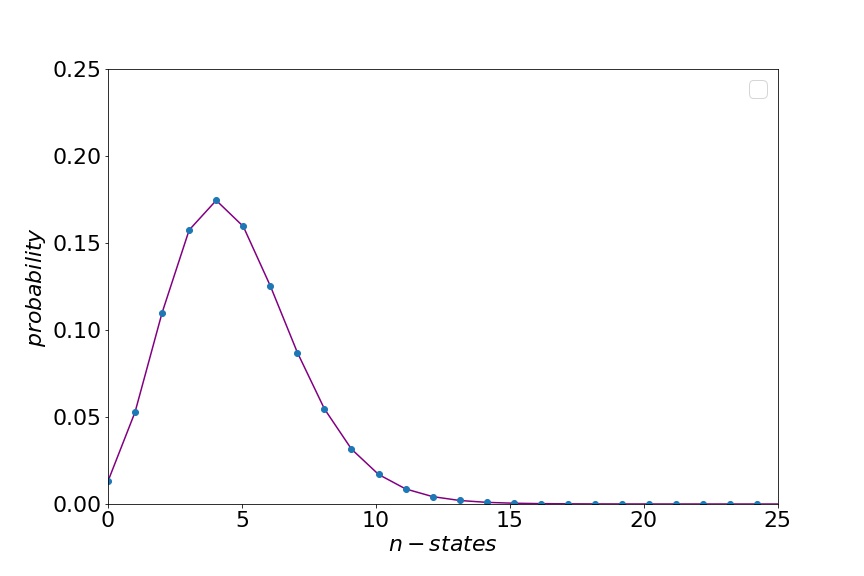}

  \caption[\large{Chapter 3: The probabilities of finding particle in new states.}]{{ Probability of finding the proton in each of the first 25 eigenstates of the non-adiabatic potential Eq. \eqref{ch3eq9} for a proton initially in the ground state. Probabilities peak around the 5th state and decrease for higher states.
    }}  
  \label{ch3f6} 
 \end{figure}
\begin{figure}[h!]
  \centering
  \includegraphics[width=1\textwidth]{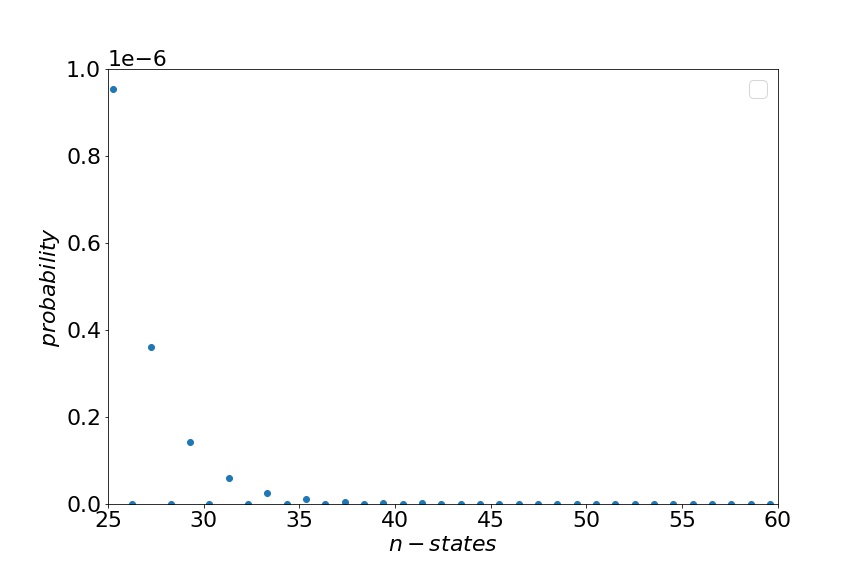}               
  \caption{ Probability of finding the proton on the right-hand side of the non-adiabatic potential well in eigenstates  $E\geq E_{25}$ Only certain higher-energy states  contribute to proton transfer relevant for mutations.
    }
  \label{ch3f7} 
 \end{figure}

\begin{figure}[h!]
  \centering
  \includegraphics[width=1\textwidth]{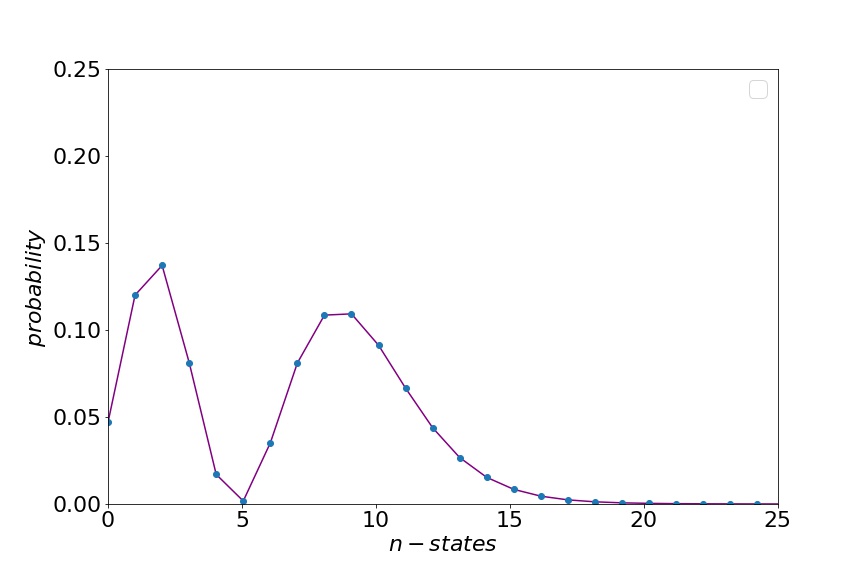}               
  \caption[{Chapter 3: The probabilities of finding a particle in each state of the new potential}]{{The probability of finding the initially excited proton in each of the first 25  $n\leq E_{25}$ eigenstates for the non-adiabatic process.}}  
  \label{ch3f8} 
 \end{figure}

 
For a mutation to occur, the proton must occupy the right-hand side of the potential well (see Fig.~\ref{ch3f1}). Therefore, we also calculated the probability of the proton appearing in eigenstates higher than the $25^{\rm th}$. In this calculation, the integral in Eq.~\eqref{ch4eq6} was evaluated from $\zeta \approx 0$ to $\zeta \approx 1.5$. The resulting probabilities, shown in Fig.~\ref{ch3f6}, indicate that beyond the $25^{\rm th}$ eigenstate, the probability drops rapidly, and Fig.~\ref{ch3f7} shows the probabilities of finding the proton on the right side of the well in eigenstates with $E \geq E_{25}$. Only a few of the energy states contribute significantly to proton transfer. Moreover, only protons occupying odd-numbered eigenstates above the $25^{\rm th}$ appear on the right-hand side of the potential well, highlighting the selective nature of nonadiabatic transitions in promoting mutations.

The sum of all probabilities of finding the proton with energy $E\geq E_{25}$ on the right-hand side of the well was calculated using the following summation:
 \begin{equation}\label{ch4eq5}
\sum_{n=25}^{\infty} (\phi_n^*\psi_0 )^2 \
 \end{equation}

for odd n's and it was found to be equal to $1.560\times 10^{-6}$.

In the copying process of $10^4-10^6$ bases by DNA polymerase, there occurs only one error\cite{Gout2013}. In the present model, approximately one error can occur in $10^{6}$ bases.

Temperature affects mutation rates \cite{Waldvogel2021}, which corresponds to an increased probability that the proton occupies an excited state.If the proton is initially excited, the result will be different. The probabilities for the proton that is initially in the first excited state $\psi_1$ were calculated. Figure 8 shows the probabilities for the initially excited proton in each of the first 25 eigenstates for the non-adiabatic process. 

\begin{figure}[h!]
  \centering
  \includegraphics[width=1\textwidth]{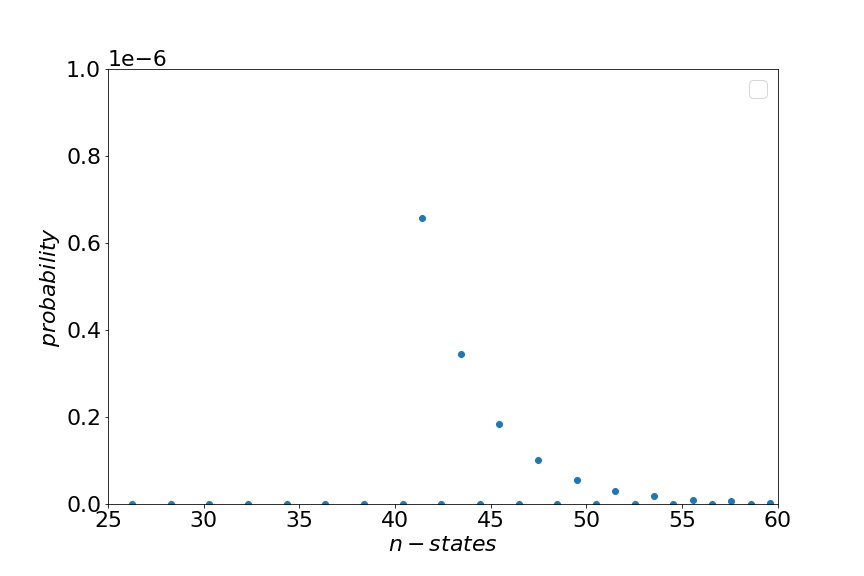}               
  \caption[{Chapter 3: The probabilities of finding an excited particle in the right part of the new potential in new states.}]{{The probability of finding the initially excited at the right-hand side of the potential well for  $n\geq n_{25}$ in the non-adiabatic process.
    }}  
  \label{ch3f9} 
 \end{figure}

 It is clear from Figure \ref{ch3f8} that from the third to the fifth eigenstate, there is a decrease in probability. Then, the probability increases for higher eigenstates up to the 9th, after which it decreases, becoming very small for the 20th state and beyond. In addition, the probabilities of finding the proton on the right-hand side of the potential were calculated. The results reveal that the transition to the right-hand side occurs after the 41st eigenstate, as shown in Fig.\ref{ch3f9}.

 The total probability was also calculated, and found to increase to 0.000534 ($\approx 10^{-4}$), compared to $1.56\times 10^{-6}$ for a proton initially in the $\psi_{0}$ state.
 
 This result is more consistent with the experimental results \cite{Gout2013} compared to previous theoretical results ($\approx 10^{-9}$)by using other numerical methods \cite{Godbeer2015}.Our model predicts mutation probabilities of order $10^{-6}$, compared to tunneling-only estimates of $10^{-9}$. Data from the 1000 Genomes Project show that over 99.5\% of mutations are point mutations \cite{1000Genomes2015}, consistent with our analysis.

\section{Conclusion}
The quantum aspect of the DNA mutation was studied. An analytical function for the nucleotide potential was introduced, which aligns with an empirical and numerical potential obtained by other researchers. The proposed potential has the mathematical form given in Eq.\eqref{ch3eq5} and represents an asymmetric double-well potential.\\
During DNA replication, the Adenine (A) and Thymine(T) nucleotides are typically connected via hydrogen bonds. A mutation can occur when a proton is transferred to the right-hand well of the potential, leading to incorrect pairing between Guanine and Thymine. 
The mechanism by which the proton is transferred to the right-hand side remains unclear, and several hypotheses have been proposed. In the present work, a newly modified model is introduced. In this model, the potential barrier that separates the two wells suddenly vanishes, allowing the particle to move freely within a single potential well. This process is non-adiabatic, occurring within a very short time and involving minimal interaction with the external environment. Given the height of the original energy barrier, such a sudden collapse of the double-well structure may be triggered by a collision with an infrared photon. ‌However, the main mechanism comes from the Heisenberg uncertainty principle. Finally, two cases were considered:
$(i)$ the proton is initially in the ground state, and $(ii)$ the proton is initially in the first excited state. 
The probabilities of finding the proton in the right-hand region of the potential well were calculated. The results show that this probability is on the order of $10^{-6}$ for the ground state and $10^{-4}$ for the first excited state. These values are consistent with the range of $10^{-4}$–$10^{-6}$ reported by other researchers.

\end{document}